# Young astronomer in Denmark  1946 to 1958

Erik Høg  - Niels Bohr Institute, Copenhagen - ehoeg@hotmail.dk

**Abstract:** This is a personal account of how I became an astronomer. Fascinated by the stars and planets in the dark sky over Lolland, an island 100 km south of Copenhagen, the interest in astronomy was growing. Encouraged by my teachers, I polished mirrors and built telescopes with generous help from the local blacksmith and I observed light curves of variable stars. Studies at the Copenhagen University from 1950 gradually led me deeper into astronomy, especially astrometry (the astronomy of positions), guided by professor Bengt Strömgren and my mentor dr. phil. Peter Naur. I was lucky to take part in the buildup of the new observatory at Brorfelde during the first difficult years and the ideas I gathered there have contributed to the two astrometry satellites Hipparcos and Gaia launched by the European Space Agency (ESA) in respectively 1989 and 2013. – The account contains notes about Danish astronomers and the observatory, but it is not meant to be a history of Danish astronomy; it is my personal memoirs.

## 1 On the island Lolland

Born in 1932, I lived in the 1930s and 40s in a rural area typical for the province of Denmark in those years. I grew up among farmers and craftsmen, the latter may have shaped my interest for instrument building. The blacksmith built my first telescope and its parallactic mounting with me standing behind, and I ground and polished my own mirrors in my parents' party room. There was a dark sky and of course no street lights to disturb my observations. I enjoyed the best Danish school education completed in 1950, but although I liked the school I was longing for the time after.

### 1.1  Family and numbers

On a day in April 1939 I walked to the village school in Frejlev on Lolland. We were three children walking the two kilometres, Ruth, our neighbour's daughter who was two weeks elder than me, and Vagn Nielsen, a five year elder boy from a few houses further up the road along the forest. Vagn had taught some of us to read and he came with us on this first day of school to help us get good seats in the class. We were very early, arriving before any body else so we easily got the best seats, just in front of the teacher. Later on, I understood that many children prefer more modest seats in their class. I could read fluently before I started in school. One of the other boys said to his mother while I was reading: Erik is reading without pointing at the line with his finger.

    Ruth and I began in school three months before we were seven. On the first day we, a dozen children, were asked one by one how far we could count. Some could count to ten, some to one hundred. I said that I could count to a million but that would take too much time, and in a way it could go on for ever because I could always say the next following number. I do not know what the teacher answered, if anything, but my capabilities were always appreciated by my teachers, except a few times. I am not saying that I am brilliant in mathematics, only rather good. At university I met people who were brilliant and later I have worked with several such people. It is astounding and a great pleasure to witness what such a person can accomplish as I have told about elsewhere.

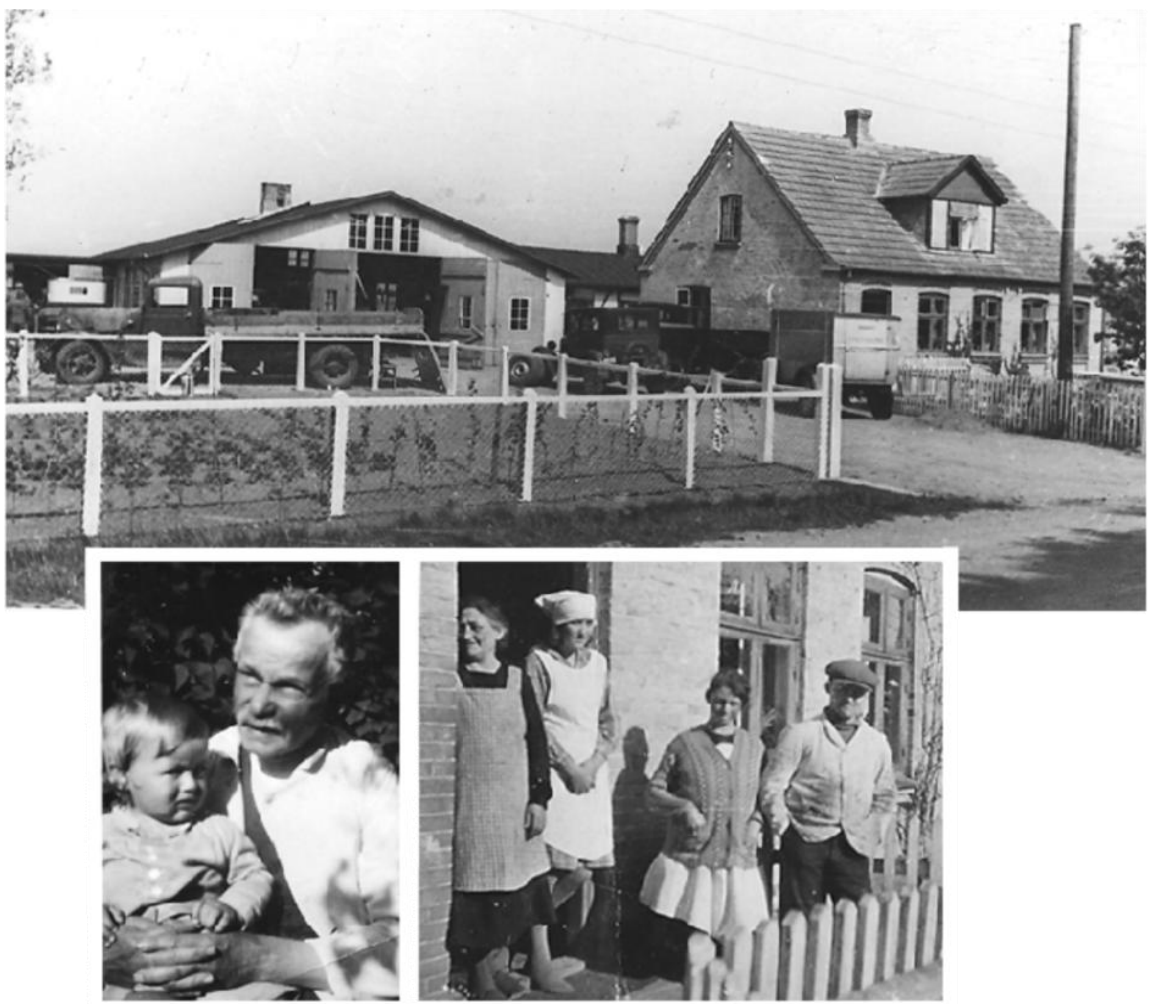

Figure 1. *Upper*: My father's workshop and house. Cars of all sizes are seen lining up for painting about 1946, just after the second World War. Old paint was removed by grinding before the spray painting. He also painted houses in the village of Frejlev in Lolland. *Lower left*: Little me on the knees of my blind grandpa who lived with us until he died in 1958. *Lower right*: Grandma, a maid, my mother and father, Agnes and Aksel Høg, about 1934. Grandpa was also a painter and built the house when he married about 1900. Behind the house were outhouses for poultry, one or two pigs, firewood, laundry, and the old painter workshop. – Photos: Erik Høg

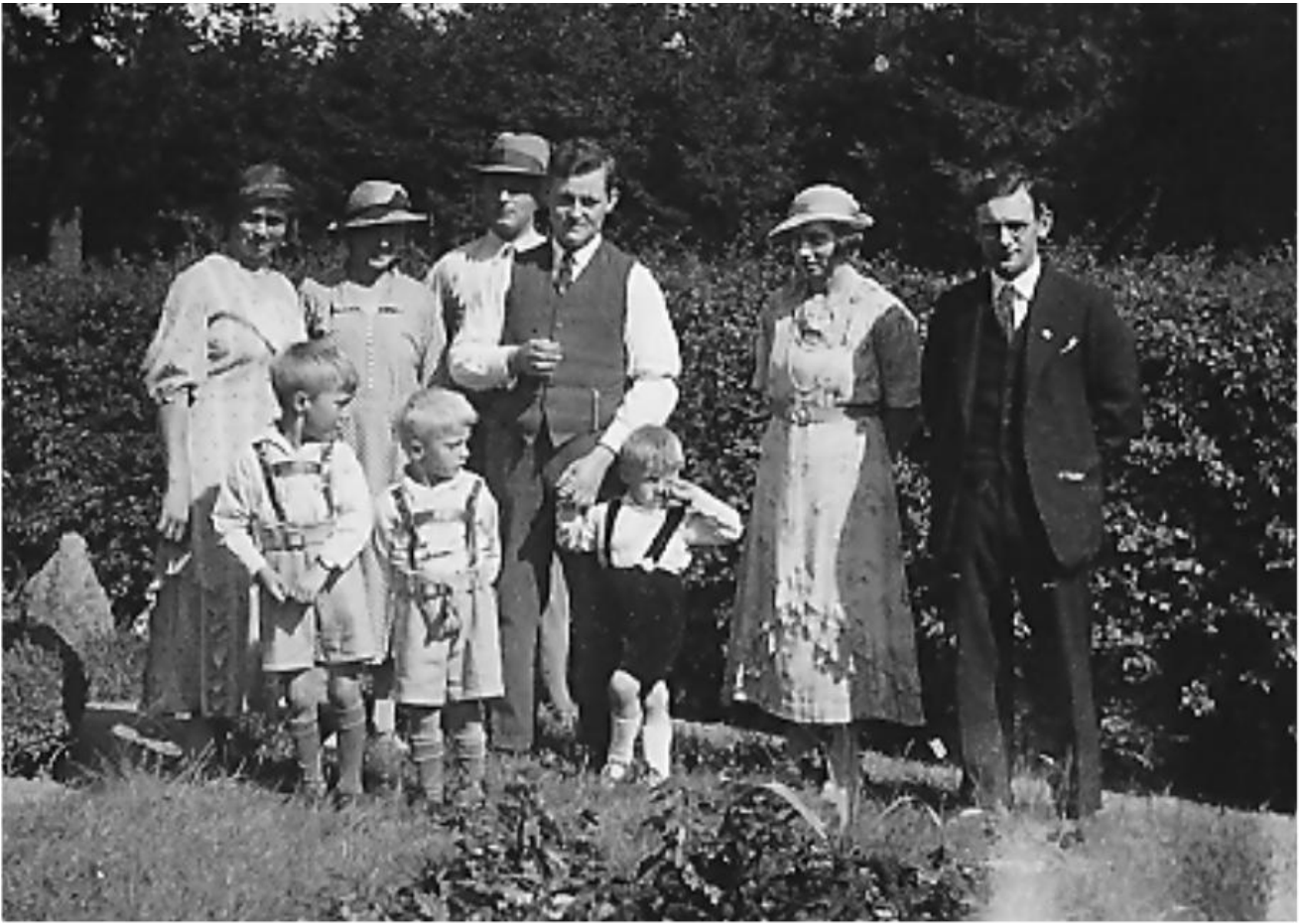

Figure 2. Three families in Sunday clothes in 1936. The three men, from left: Helge Jensen, my father Aksel Høg, and Asger Dresen, lived all their life in the same houses at the one kilometer road along the forest in Frejlev. Helge and Andrea's sons Jørgen and Poul are looking at the sulking Erik whose playmates they soon became. The ladies, from left: Agnes my mother, Andrea, and Klara, the sister of my mother. - Photo: Erik Høg

Born on 17 June 1932 as the eldest son of Agnes and Aksel Høg I grew up with my four year younger sister Karen and eleven year younger brother Jens. Our mother was always there to care for us and she lived to the biblical age of 93, quite healthy all her life and eager and able to talk with us until her last hours. Our father was a house painter as his father had been. I have enjoyed living in social environments with great stability.

Hit by his last heart attack my father died on 19 July 1968, just three months before he would have stopped working as he had done all his life taking every year perhaps one week vacation with no income. He could then have received the people pension which every Danish citizen would get with 67 in those times.

Mother phoned me in Hamburg, Germany, where we lived those years. She told me what had happened and Aase, my wife, only saw my face turn pale white while I said nothing, just standing with the phone listening to my mother. Aase has only seen that change in my face once again, five years later. We agreed that I should take the train to Denmark immediately and Aase would come the next day by car with our two children.

My parents never thought that I should take over the painter business which was too small any way, and they wanted I should have an education, encouraged by the teacher Mr Winter Andersen in the village school. My mother had wanted to become a medical doctor, but she always had headache and the money was not there. When she was young her father and others in the village had lost a lot on gold mines in South Africa.

Father had a talent for music although I never heard him sing. When young he had learnt to play the violin and the transverse flute, he even took lessons in Copenhagen. Mother told me that after their marriage he only played the flute for her when she had born a child. We have several oil paintings from his hand and also from my grandfather, sometimes enlarged versions of a motif on postcards which I later found. My sister Karen has cultivated her talent as painter on aquarels and porcelain.

In summer time in the 1940s my uncle and his two children came from Copenhagen to spend the two weeks vacation with us in our rather big house. Mother had to care for everything and one evening she burst into tears and complained, but she was a strong woman and next morning she cooked and served for us as usual. House and garden was definitely women's job in those years – I am enjoying that this has changed.

Our parents held the traditional celebrations at, e.g., confirmation of the Christian baptism of one of us children at the age of 14 or their own silver wedding, i.e. 25 years of marriage. Tables for some 50 guests were arranged in the painter workshop, food was served in late afternoon, cooked by hired women in the kitchen. Eating and talking was interrupted by speeches of several speakers with much experience from similar occasions, among them the parson and the school teacher who were invited with their wives. Toasts with lifted glasses and loud Hurras for the celebrated person had to follow a speech. Often a song from a traditional book or one written for the occasion by a local poet would follow. Aase is good at making that kind of poetry while I have no such talent. The dinner could last four hours before the tables were cleared. Coffee and home baked cakes were offered, cigarettes and big cigars were smoked. Groups had then found a table where cards were played, brought along in nice boxes. My brother Jens was good at l'Hombre already as a boy, I believe he prefers Bridge nowadays. I was never very good and never like to play cards except with our grandchildren who expect it, but Aase luckily likes to play. Sometimes a local orchestra played for dance at the party for a few hours. Everything ended shortly after midnight when a "go home meal" of soup and bread was served.

**1.2 School and astronomy**

All education in school and university was and still is free of charges in Denmark. Today, all students at university or higher education receive a public subsistence support, in my time only few received and I was one of them. When I was eleven I should go to the middle school in the Katedralskole, the grammar school of the town Nykøbing on the island Falster, but I first had to pass a test together with many other children. During the months before, I was prepared by the two teachers of the village school. They drilled me for an hour after the ordinary school time the three days of week we came to the school, and I got extra work to do at home.

One day, the teacher Winter Andersen was standing behind me looking out of the class window when he asked: Erik, what will you do if you do not pass the test for the middle school. I turned my head to see him and immediately answered: I will never try again! That was my nature then and I am sure the teacher has smiled inside because he knew I would pass the test.

I was the first from the village going to the town school. Quite soon a few other children followed, mostly boys, but it was not until some twenty years later that the majority of children in my part of Denmark received more than seven years of basic school education. Nearly all my seven years in the town school I did the 8+8 km six days a week on bicycle in wind, rain or snow. That is why you keep such a good health, said a teacher. Only the first three years did I take the local train during the three winter months which meant that I had to leave home in the morning about half past five, going on bicycle to the train station. Some older boys from the school sometimes mobbed me when we drove home. I was not allowed to speak, I then wrote to them on a piece of paper. I was not allowed to sit on a seat, but was placed in the luggage net above the seats, which the ticket controller forbade. They once kept my cap before I got off the train which almost made me weep, but they threw it out the window when the train had gone a few hundred meters. When I told my mother she phoned one of the parents, the local parson with whom we always had very good relations.

I did however sometimes fight other boys in the middle school. Once I swung my right fist towards someone's ear but he put his bag between and I hit the hard bag instead. This was good for him but my hand began to hurt and swell. Mother said that I should visit Mrs Clausen the "clever woman" in our village since she had helped so many others even coming to her from far away. The lady massaged my arm, but the pain continued and finally, six weeks after the fight with the boy I visited a real doctor in the town. He saw that a bone in the hand (metacarpus) was broken and ordered me not to use the hand at all for a few weeks. That helped but for many decades after a bump on my right hand was still visible. Since then I sometimes tell about the "clever woman" who massaged my hand without realizing it was broken.

After school time I often visited the central library in the town where the shelves with mathematics, physics, astronomy and chemistry had my interest, but I also read Nordic sagas, the Bible, the Koran, and Dante's Divine Comedy, though only the first part: Inferno (Hell). In school I learnt to appreciate the classical literature, but I did not read much literature in my mature life. When about sixty, however, I decided to read these books, or at least to look at them again, before I would become too old to understand and enjoy them.

What drives me is to understand the big things, the big issues: the universe, the nature on Earth, human beings, the individual person. What he or she thinks about life, about God and religion. The belief of the individual person interests me even more than the official religions. I have found that most believers have their own personal religion, different from the official, and many have realized that God and all gods are created by man, in the heads of men and women.

What drives me in astronomy is to do something useful for science. It turned out that I had a particular talent for invention of new instrumentation, for its implementation and its use in observations and finally to produce a good catalogue of stars, all this in collaboration with a few or many people as need was. These catalogues have been used by thousands of people and I feel that I have done well.

You must love your daily work in order to accomplish something in science and elsewhere, it is not enough to love the stars or the knowledge about the universe. You must live every day as if it is the most important day of your life. Not the last day of your life, but the most important. You must do your best every day, and luckily I have had the energy to do so and a family and environment where it was possible and appreciated.

Being the first in our family to visit a grammar school has raised some thoughts and emotions in our surroundings most of which we could only guess about. My mother was careful where to speak about me and my achievements, and she once warned me about this problem with a part of the family who could be very outspoken. My father anyway spoke very little and he was a very kind person while my mother could be rather strict and quite outspoken.

The question came often forward: "What is the practical use of it?" I temporarily convinced the men at a lunch party at home how useful my chemical experiments could be when I made a mix to lower the temperature of their bottle of snaps by over ten degrees; it was NOT ice and salt of course because we had no ice. The strong Danish snaps must be cold when indulged to herring which was impossible at summertime before the days of the fridge. My chemical experiments to make gunpowder and fireworks were also quite successful but not so appreciated. My father discussed the safety issues with me and my playmate Kjeld and reasoned us, not ordered us, to dig a hole for the stuff in the forest. In the 1940s we disposed of rubbish either by burning in the kitchen range or by tipping in the forest or in a hole in the moor from which my father with others had dug peat during the war. These rubbish tips later became illegal and some of them were then emptied.

A few specific remarks about my future I still remember. Two of my uncles lived near to my parent's house and we saw them often. They both had a small apple plantation at their house and most of their life they worked at a large plantation nearby. My uncle Asger repeatedly urged me to become a dentist: "They earn a lot of money". The other uncle, Oskar, once said that he would not count my good head unless I would be able once to own a villa at Strandvejen, a very wealthy neighbourhood north of Copenhagen. But Oskar always asked me to tell something when I had been abroad, he was one of the few who did so, and I liked to tell. A third uncle said to my mother: "What a pity that Erik has such a strange interest as astronomy when he is so bright." All this happened when I was a boy but I liked all three uncles throughout their lives and they always enjoyed to see me and my family.

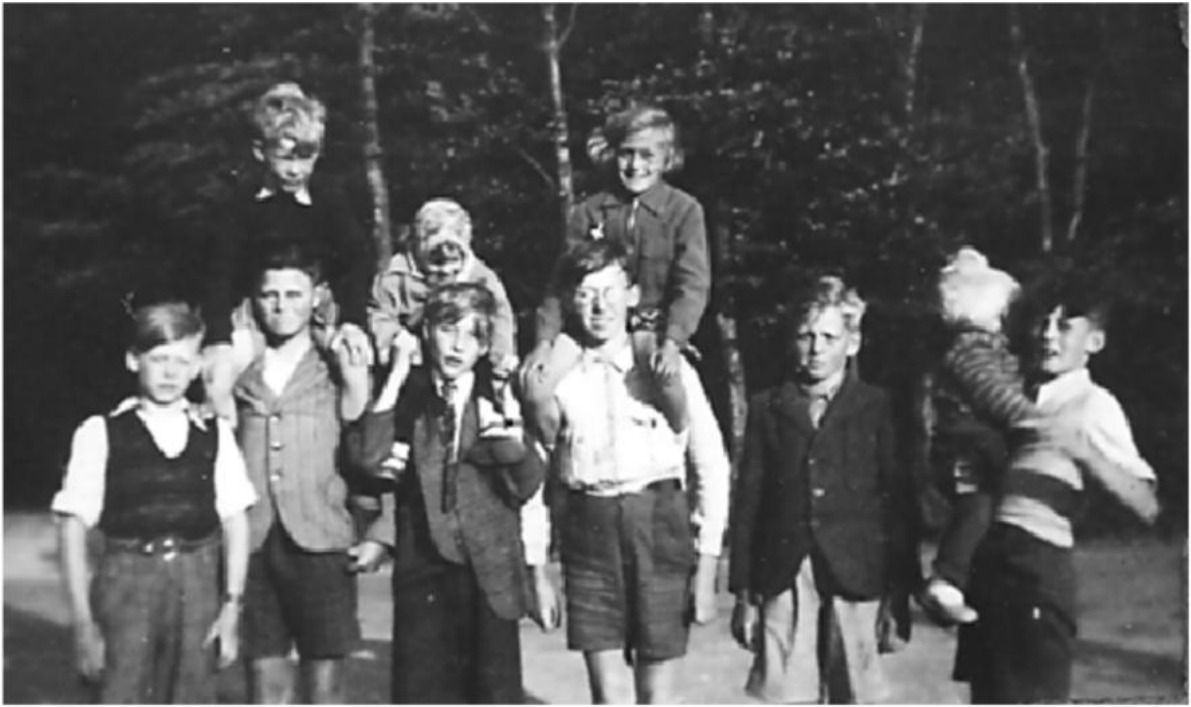

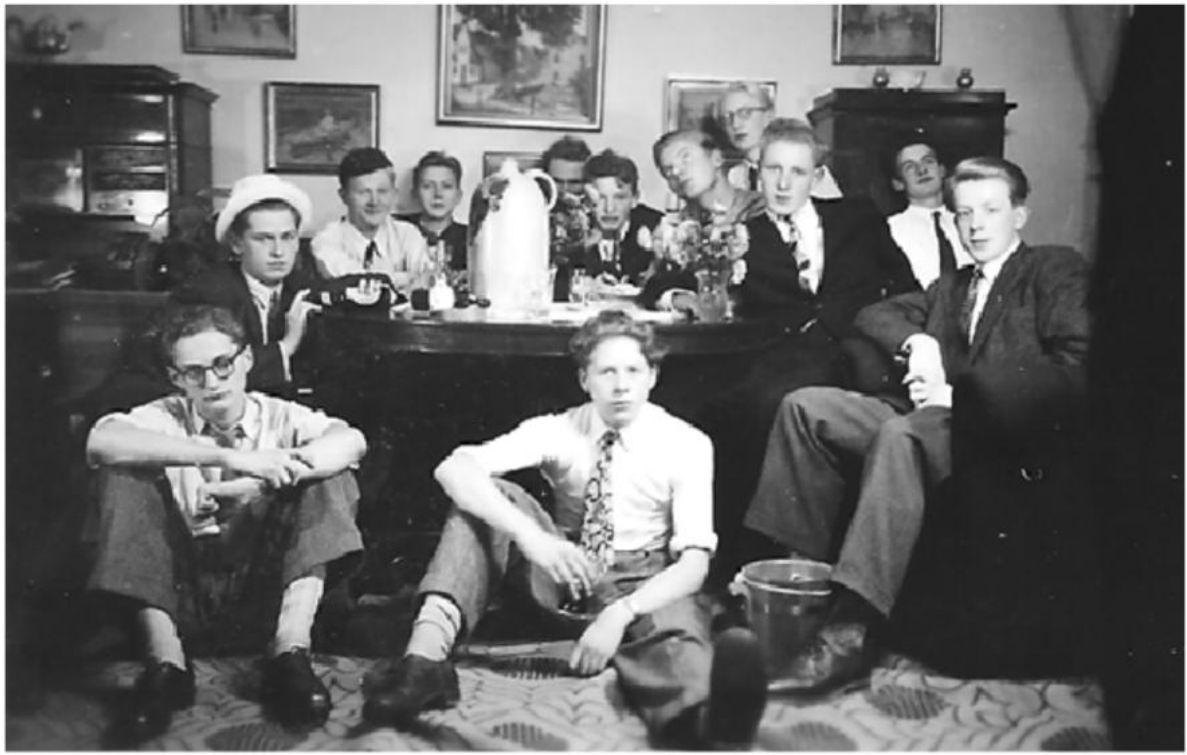

Figure 3. *Upper*: Children at our house on my birthday in 1945. I am standing at the right with my little brother Jens and my sister Karen is high on shoulders in the middle. We played much in the forest behind. *Lower*: Party of school boys 1949, rather late in the evening it seems. I am sitting far back at the right. – Photos: Erik Høg

## 1.3  The stars and my three telescopes

My interest in astronomy probably started while reading a few rather old popular books from my grandfather's small bookshelf. The interest led me to build a telescope of eyeglasses bought in the town shop when I was fourteen. Small astronomical telescopes have since become available at a reasonable price, but for me out of a family of craftsmen it was natural to make things myself, my fingers were built for that and I had the patience. I understood elementary optics and just said which strength of glass, i.e. which focal length I needed for the telescope.

The popular journal Nordisk Astronomisk Tidsskrift, issued at the Copenhagen Observatory since 1920, I found in the school library and I read all issues. C. Luplau Janssen's "Stjernehimlen og dens vidundere" from 1938 with its engaged style suited me, although I could later see that it was quite old fashioned. I bought a Finnish book because it contained tables about the planets, and therefore started to learn Finnish. I could then read the headings of the tables but not much more.

I bought Amateur Telescope Making by Ingalls and Porter in order to grind mirrors, first 8 cm diameter, later 12 cm. The grinding was done by very hard grains of Carborundum with water between the mirror and another glass plate below. The lower plate was fixed to the top of a big barrel put in vertical position in the party room of our house.

Pushing the mirror back and forth for hours, days, and weeks makes it hollow, concave, of the desired curvature corresponding in my case to a sphere of two meter radius. This would mean the light from stars would be concentrated at the focal plane, one meter from the reflecting mirror. After sufficient grinding with ever finer grains of Carborundum (silicon carbide) the mirror is ready for polishing. I had acquired ten grades of Carborundum from a shop in Nykøbing. The polishing is done on black pitch melted on the lower glass plate and covered with a red powder, also on top of the barrel and again pushing back and forth. A mirror must be tested for its perfect spherical form during the polishing. I set up the testing of the mirrors on the long party table. I used the Foucault method with a razor blade as edge at the reflected image of a pinhole and also the Ronchi method where a grid is used.

My physics teacher, Mr Kappelgaard, and the head of school, Mr Willesen, sometimes took our road on their Sunday tour on bicycle. Once they stopped at our house and I ran out to meet them. Kappelgaard joined me in to see my setup, I said that I really needed another cleaner room for the polishing, otherwise the mirror could be scratched from grains of the hard Carborundum used for grinding, as I had read in the book. He calmed me and said I should be satisfied with conditions as they were.

When the polishing is done the glass surface reflects about five per cent of the light. This is sufficient during the testing but not for observation of stars. The surface was therefore coated with silver deposited from a solution according to a well known recipe, but not always with satisfactory result. The silver surface reflects nearly hundred per cent if it is perfect, but it darkens especially by traces of sulphurous vapour in the air. The mirror must then be cleaned and coated with silver again. Professional astronomers and amateurs nowadays usually use mirrors coated with aluminium in a vacuum chamber and this surface holds good reflectivity much better than silver. However, much later when I became involved with the satellite Hipparcos I learnt that silver deposited in vacuum was preferred because it has higher reflectivity than aluminium. Such a silver surface does not deteriorate when flying in space because of the natural vacuum there.

One of the village blacksmiths, Mads Peter Rasmussen, built the mechanical parts of a parallactic mounting for the 8 cm telescope. With such a mounting the telescope can be easily pointed to any star in the sky by rotation about two axes. The polar axis is parallel to the Earth axis and the other one perpendicular to it holds the telescope. I was standing there instructing the old smith what to be done. He hardly said anything, but I am sure now that he enjoyed having a boy telling him to build an astronomical telescope. The tube for the telescope was built by another smith in the village, Jens Rasmussen, who only asked a symbolic pay of 12 kroner (= 2 dollars) on 30 Dec. 1947, according to the diary kept during my school days.

The mechanics for the 12 cm telescope was built in a workshop in the town Nykøbing since the village blacksmiths could not do it. It was later sold to another boy in the village, Poul Henning, son of the shopkeeper, who used it for many years. I sold it because at that time I had moved to Copenhagen more than 100 km from Frejlev so that the telescope was of no use for me anymore. Poul Henning came to me sixty years later and told that he still had that telescope and that I should now have it back.

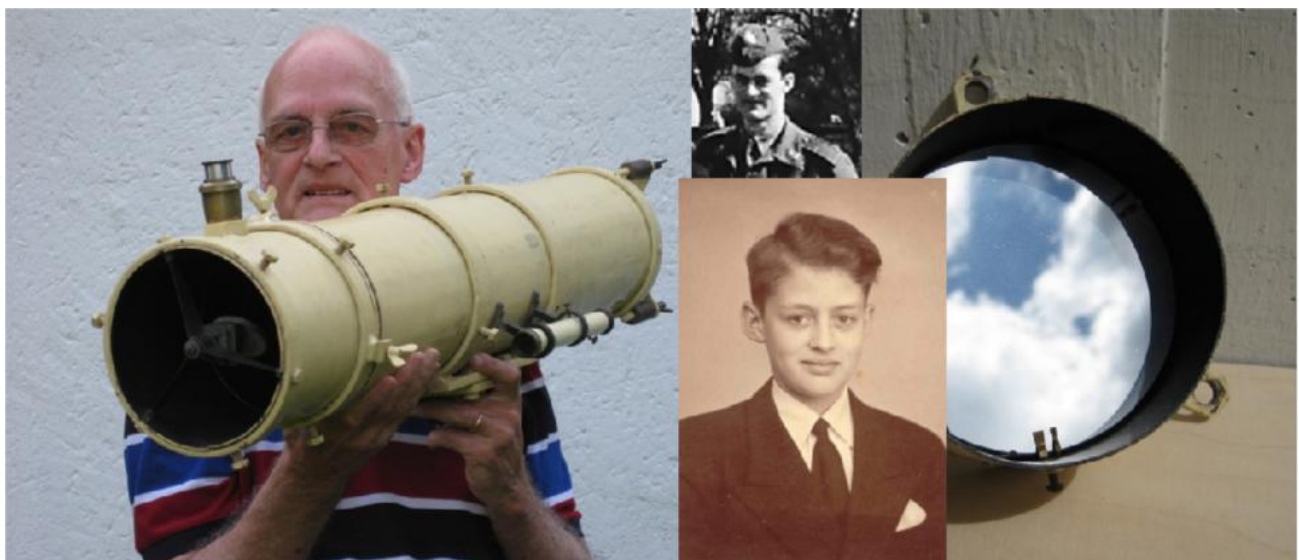

Figure 4. My third telescope with 12 cm aperture was returned to me in 2010. - From I was fifteen I ground and polished mirrors and observed variable stars. After moving to study in Copenhagen in 1950, I sold the telescope to a boy in the village. He came to me 60 years later and gave it back, it now stands in our living room. **Right**: The silver was deposited chemically in 1949 and is obviously still untarnished, reflecting clouds and the blue sky. **Inserts**: The upper photo is from 1956 when I was a soldier, the one below is from 1946 when fourteen. – Photos: Erik Høg

I observed variable stars as organized by the Danish Astronomical Society. A variable star changes its brightness in the course of time and I recorded the estimated brightness, which is called magnitude in astronomy, night after night in a nice protocol. The magnitude is estimated without any special instrument, just by comparison with other stars in the field of view using a map of the field received from the Society. This method is called the Argelander step method after the famous German astronomer of Bonn who developed the method around 1840 before any instruments, i.e. photometers, had been developed for measurement of magnitudes. Many years later, in 1999, I had the honour to give the first Argelander lecture at the Bonn Observatory.

But all my magnitude estimates of variable stars have been lost because my dear wife found the book and tore out those pages, so that she could better use the blank pages for her notes on good cooking. When I realized much later what she had done I could do nothing but laugh a bit and forgive her, although I have always liked to keep my old notes.

A girl in my class reminded me thirty years later that I left a school party rather early because I had to go home to observe a star. In fact I can still see the clear dark sky over me with all the stars while I cycled the eight km home on that night. It was the variable star SS Cygni, a dwarf nova in the constellation Swan, which I had never seen before because it was too dark for my telescope most of the time. But it sometimes brightens and the task for the observer is to record when this happens. I had a small finding chart with the stars at the spot of the sky so that I could recognize it. I did catch the nova that night when it brightened, which meant quite some luck because that happens at intervals varying between four and seven weeks.

While observing another night I was overwhelmed by the sight of a meteor shower which I had not even heard of before. It must have been the Perseid shower which comes in August. The sky blinked and blinked with hundreds of shooting stars, each of them being caused by a dust grain, a meteor. The grains belong to the comet Swift-Tuttle and have been left behind in the comet orbit in the course of time. Every year the earth rushes through the crossing point with the comet orbit and the high atmosphere hits the grains at a speed of 40 km per second with the result that the thin air shines. Thus the light comes from the atoms of air and not from the glowing meteor which is too small to be seen. The Perseid meteor shower has been observed for about 2000 years, with the earliest information on this meteor shower coming from the Far East. Some Catholics refer to the Perseids as the "tears of St. Lawrence", since August 10 is the date of that saint's martyrdom.

When I was about fourteen years I built an electrically regulated pendulum. It was not very reliable and not impressive at all, but I liked to do such things. I filled a notebook with calculations of the number *pi* with many decimals, following a method found in an old book by the Danish author Poul la Cour.

My teachers of physics and mathematics encouraged me in these activities. I rather liked to go to school and did all work quite carefully, but I was longing for a time to come when I should only work with my real interests; no more history and languages. Nevertheless, I have since read a lot of history and learnt a useful amount of several languages. The last languages were Bahasa Indonesia, Japanese and Chinese which I learnt to speak a bit, the last two not to write, when I was about fifty. I have always been careful to learn a good pronunciation of a language from the very beginning. Ever since, it gives great fun to take a chance to practice and to see the smile when he or she unexpectedly hears the mother tongue be it Chinese or Japanese.

While I prepared for the final exam, the "studentereksamen", in 1950 I amused myself learning shorthand writing. I could write notes quite nicely, but never gained any high speed. All we sixteen students passed the exam and we celebrated our completion of school at parties arranged by our parents during many days.

Our celebrations led to a scandal in town because some of us had drunk too much at the first lunch invited by one of the parents. The same afternoon we went to the garden of the head of school, L. Willesen, to listen to his speech and to have the traditional photo taken of the group of new students. But I could only walk the one kilometre to the garden because two of my classmates almost carried me by arms and some of us could not keep the row while the photo was taken. During Willesen's speech I saw, and I can still see, the glass of port in my hand slowly tilting and its content flowing towards Gerda's white gown. My inner film stops here, but I know that many saw it happen and that nobody screamed or shouted at me and I never heard a real blame for it, only amused laughter on the top boy of the school. I have always been surrounded by kind and forgiving people – or nearly always. Mrs Willesen kindly helped Gerda.

But it could not pass unnoticed and Willesen wrote to the parents blaming them for having let some of the young people drink so much that "they could not control themselves." We students continued our celebrations at a great dinner at other parents where I tasted champagne for the first time in my life.

When we heard next morning of the letter from Willesen the three of us who had shown the least control of ourselves went to see him in his office. We explained that not the parents were to be blamed, it was our fault. He listened without saying much, but at the great celebration in the town theatre a few days later where the whole school and all parents were present Willesen mentioned the incidence in his speech. That brought him great blame in the local newspaper: He should not spoil the great days for the happy young people neither for their parents.

My classmate Thomsen has said with some satisfaction: "We were a bit ahead of our time in Willesen's garden." A youngster today would probably shrug his shoulder and wonder: "Was that really enough in those days to be called a scandal?" In fact, since long most of us gather every year in June to a small party including our partners. In 1950, however, the Korean War began on 25 June, a few days after the final examination and we listened to the news in radio in between our happy parties.

## 2 Danish astronomy in the 1950s

> The main issue of the 1950s in Danish astronomy was the development of a new observatory site 50 km outside Copenhagen. The Copenhagen University Observatory was originally situated on top of the Round Tower at the centre of the city. The tower was built 1637-42 and its first director was the Danish astronomer Christian Longomontanus. In 1861, the observatory had been moved to a new building on the previous rampart of the city, a good site for an observatory at that time, but no longer satisfactory in the 20$^{th}$ century.

### 2.1 Planning a new observatory

The history of Danish astronomy has been treated by Nielsen (1962), Moesgaard et al. (1983), Gyldenkerne (1986) and Gyldenkerne et al. (1990), all four in Danish, see comments in Sect. 6.3.

Seven pages in the last volume contain the only known account of this history in English and since it is covering 400 years, less than one page is about the 1950s. Accounts of the recent history are given by Andersen (2004) and Petersen (2015). The book by Moustgaard (1990) is a thorough historical study, also in Danish, of the years 1900-1950.

Professor of astronomy in Copenhagen was Bengt Strömgren (1908-87). He was for a long time the only professor of astronomy in the kingdom of Denmark, but he was staying in the USA, first as director of Yerkes and McDonald Observatories 1951-57. In 1957, he was appointed the first professor of theoretical astrophysics at the Institute for Advanced Study in Princeton, where he got Albert Einstein's office. He stayed at Princeton with his family until 1967, when he went back to Denmark, and became the second-last resident of the Carlsberg Mansion of Honour, which had earlier been occupied by Niels Bohr among others. In 1987, he died after a short period of illness.

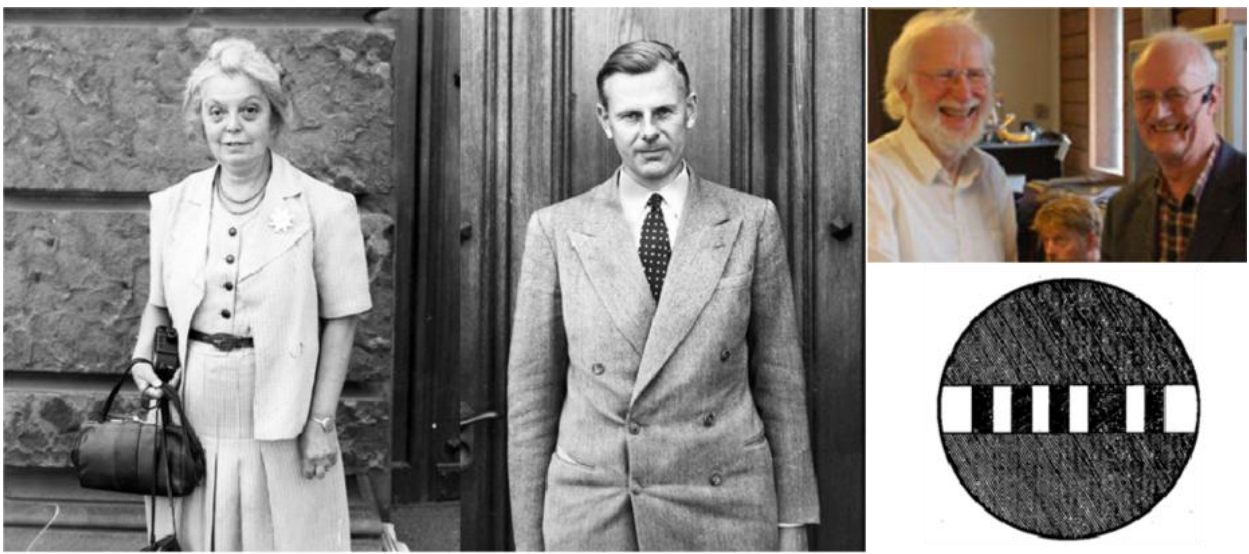

Figure 5. *Left*: Julie Vinter Hansen and Bengt Strömgren in 1948. *Right*: Peter Naur and Erik Høg in 2010. The slit system was used in 1925 by Bengt Strömgren for experiments with photoelectric recording of transits. A photocell behind the slits gave a signal in which the transit time for each slit could be detected thus obtaining the right ascension of the star. The slit system was placed in the focal plane of the meridian circle at Østervold. This idea led me to propose photon counting astrometry in July 1960. – Photos: International Astronomical Union (IAU), Niels Bohr Institute (NBI), and Bengt Strömgren

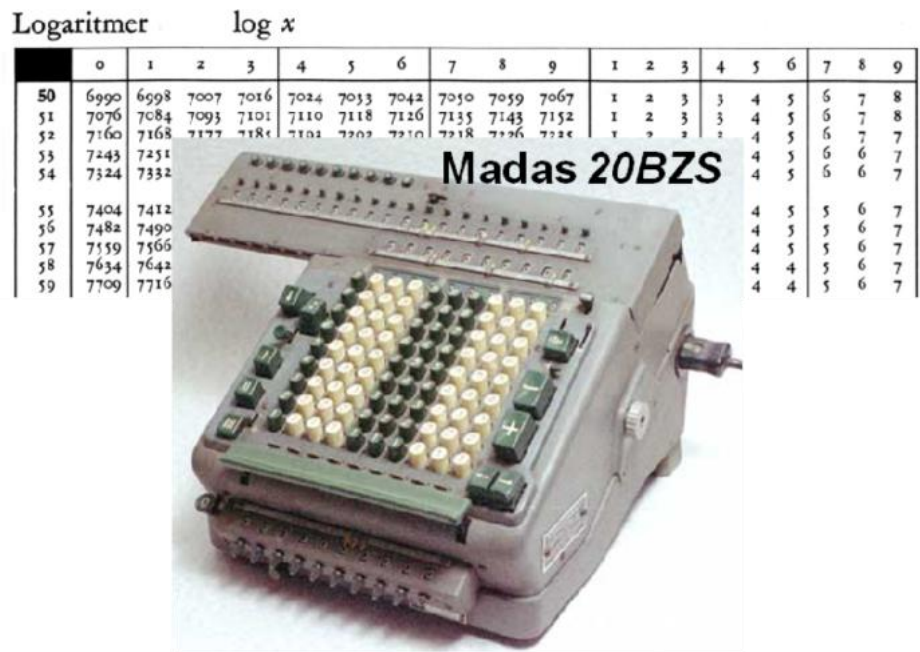

Figure 6. Calculation tools in the 1950s. – At Grammar school, multiplication and division were carried out with 4-decimal logarithms contained in a table of some eight pages including antilogarithms. At the university, a student had a book with 5-decimal logarithms. At the observatory I had a mechanical calculator and only Peter Naur had a luxurious electro-mechanical Madas with 10 digits input and up to 20 digits in the result of a multiplication.

Already in the 1940s, Bengt Strömgren had begun to establish an observatory outside Copenhagen, away from the city light, dust and vibrations, and his long absence in the USA did not stop these efforts. The main instrument was to be a new meridian circle. Strömgren, himself a brilliant astrophysicist, knew the importance of astrometry for astronomy, and the observation of positions of stars could well be obtained in our Danish climate, rather than astrophysical observations because the spectrum of a star is changed much by dust and haze in the atmosphere. In astrometry, however, the direction to the star is measured and this is not as severely changed in the Danish climate as the spectrum is.

Funding for the instrument was granted with 400 000 DKK (then about £20 000) by the Carlsberg Foundation in 1944 on the occasion of the 300 centenary of Ole Rømer (1644-1710). A place 50 km to the west of Copenhagen near the small village of Brorfelde was chosen after seeing observations had been made. In 1947 a farm was bought by the University giving an area of 39 ha around a hill with an altitude of 90 m on which the meridian circle should be built, Gyldenkerne (1986).

The meridian circle was invented by Ole Rømer and used by him up to 1710. For two centuries, from about 1800, it remained the fundamental instrument of astrometry. With this type of instrument large angles on the sky could be measured very accurately, connecting all stars into an accurate celestial system of positions. This task has been taken over by space astrometry with the ESA satellite Hipparcos observing 1989-93 and rendering meridian circles obsolete. But nothing like that was expected 50 years ago, the improvement of meridian circles was a challenging task to which Bengt Strömgren had given first priority with the Brorfelde observatory. - I was the lucky first student set to work on that task in 1953.

The instrument was ordered at the British firm Grubb Parsons in Newcastle and was delivered and erected in 1953. In the same year 1953 the old meridian circle in the observatory of San Fernando, Spain, was replaced by a twin meridian circle. This was not a result of any coordination by Bengt Strömgren as I have always guessed, but a Spanish commission visiting the Royal Greenwich Observatory in England in 1946 heard that a meridian circle had been ordered for the Copenhagen Observatory. They had come to seek advice for a replacement of their old instrument and immediately ordered a twin for the funding they already had. This is the information recently received from the curator in San Fernando, according to González (2004). It is quite amazing that the twin meridian circles were erected in the same year without any coordination between the two observatories! Much later, in the 1980s, a collaboration was begun as mentioned below.

It was felt in Copenhagen, however, that funding for Brorfelde from the university for the mechanical workshop, main building, houses for personnel and for further instruments came very late, and that may have contributed to Strömgren's extended stay in the USA.

The first astronomer, dr. phil. Peter Naur, settled in Brorfelde with his family in August 1955. His task was to make the meridian circle a working instrument, but it took nine years and employment of more staff to reach that goal. With Svend Laustsen as leader from 1959 the first useful observations of stars were made in 1964.

# 3 Student in Copenhagen at Østervold

I have often been encouraged to write my scientific biography and the following is about the 1950s. The years in Hamburg after 1958 are treated in Høg (2014a). I try to cover the persons and the environment in which I grew up as a student and young astronomer, and some letters between the leading astronomers of the time have given me unusual insights in the events.

### 3.1 My life with meridian circles

Astrometry fascinated me because I understood the importance for astronomy and I was guided into the work with meridian circles by chance and by the care of my supervisors, to a task where I had the right talents. Most of my time for the next twenty years was then dedicated to meridian circles and it brought me great satisfaction.

I recall a summer day in 1952 when I visited Kjeld Gyldenkerne (1919-1999) by bicycle from Copenhagen. He lived not far from Brorfelde and the two of us climbed the big concrete pier for the instrument to come next year and we were sitting on the top. Being a twenty-year youngster, I could see that something great was in the making, and Gyldenkerne often gave me confidence in myself, as so many others in fact had done.

My mentor Peter Naur (*1928) was however concerned about giving me that task because it meant that I had to spend much time out there completely alone and he saw the real danger that it could kill the interest in astronomy in any 21 year young man. But by some lucky coincidence it turned out that I had a talent and a liking for just such development of an instrument and this work was basis for my final exam.

Peter Naur himself was dedicated to the development of computers and to the use of modern electronics, both techniques were then very new in astronomy. I learnt from him about computers and electronics earlier than most other astronomers and this experience was crucial for my future career when I, for example, designed a satellite for astrometry in 1975, launched by ESA in 1989 with the name Hipparcos. Hipparcos gave positions, motions and distances for a large number of stars with unprecedented accuracy; for an overview of satellite astrometry see Høg (2014b).

### 3.2 Østervold observatory

Observator Julie Marie Vinter Hansen (1890-1960), Figure 5, was the daily leader of the Copenhagen and Brorfelde observatories in the years until 1958 when Anders Reiz (1915-2000) succeeded Bengt Strömgren as professor. She had very frequent contact by letter with "Bengt" as she called him, for others he was of course always "Professor Strömgren". I possess copies of several of these letters which show the very friendly tone and complete loyalty between the two.

This was the stage setting in Copenhagen during my studies starting in 1950 when I was 18. We only saw Strömgren every summer when he visited Denmark for a month or two. I then spoke with him and we often corresponded. I greatly admired him and I am deeply grateful for what I have learnt from him and his papers. He was very active in 1980 to ensure that Denmark would vote in favour of the Hipparcos mission in ESA where the competition with other space mission proposals was tough.

In January 1953 I had finished the courses in mathematics, physics, chemistry and elementary astronomy and could concentrate on astronomy. There were two to four other students of astronomy, only one of whom, Svend Laustsen (*1927), became a professional life-long astronomer. He studied astronomy during the same period as I, completing his exam also in 1956. We students saw each other mainly at the lectures given weekly, always in Danish.

Peter Naur lectured in the autumn of 1954: (1) *Structure of the Milky Way system*. Amanuensis, dr. phil. Mogens Rudkjøbing (1915-2007) lectured in four semesters in 1953-55 as follows: In a presently unknown semester: (2) *Theory of stellar atmospheres and theory of stellar interior*; in unknown semester: (3) *Stellar statistics*; in spring 1955: (4) *Kinematics and dynamic s of the Milky Way*; in autumn 1955: (5) *Main instruments of astronomy*. Rudkjøbing had his desk in the library where he worked every day. He was an astrophysicist and close collaborator of Strömgren.

It was interesting to hear recently from Jørgen Otzen Petersen that he was impressed by the number of courses in atonomy at the Observatory as reported here compared to what he recalls in the three year period 1958-60. He himself had only two courses in astronomy held by professor Reiz: stellar structure and evolution and interstellar matter, but there may have been more. Two courses in relativity theory at the institute of physics were added, and that was enough for his final exam.

Amanuensis, mag. scient. Karl August Thernöe (1911-1987) taught us computing using mechanical machines and, e.g., L.J. Comrie's book on interpolation and books on computation of orbits. Magister Thernöe, as he called himself when the phone rang, had his office in the east basement of the observatory in the room of the Time Service for Denmark which Thernöe checked every day by radio signals from Greenwich. Once in 1953 when I was staying there in the evening as so often, two persons entered unexpectedly, Julie Vinter Hansen and perhaps Rudkjøbing. I was doing something I had no permission to do and therefore did only secretly, I was using Magister Thernöe's Brunsviga machine. It was electro-mechanical and much more convenient than the purely mechanical one I was supposed to use. So I felt quite ashamed at first, but then I saw Julie's kind smile. There was no reproach in her face and I am sure she was pleased to see the keen interest of one of the very few students of astronomy at that time. She was usually called Julie among us. The most modern calculator in the observatory was the electro-mechanical Madas machine which Peter Naur was privileged to use (Figure 5).

Kaj Aage Strand (1907-2000) from the USA gave a course in astrometry in 1954. Strand was Danish by birth and became director of the US Naval Observatory in Washington DC where I often enjoyed being his guest. Lectures about my work were well received, especially since the USNO was and still is the observatory in the USA with the largest amount of work on astrometry.

There were sometimes foreign guests at Østervold giving a colloquium, but in fact no weekly seminars took place in the observatory. On 11 Dec. 53 Vinter Hansen writes that she, Rudkjøbing and Naur "will try to arrange colloquia in the next semester", but colloquia were in fact only given occasionally. Naur gave a talk (in Danish) on 16 Feb. 1954 on *"Studies in recent years of colour-magnitude diagrams"*. Jan H. Oort (1900-92), director of Leiden Observatory, gave a presentation of radio astronomy in 1957 to a large audience, introduced by Bengt Strömgren.

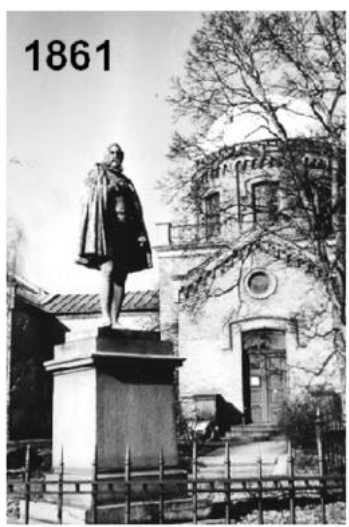 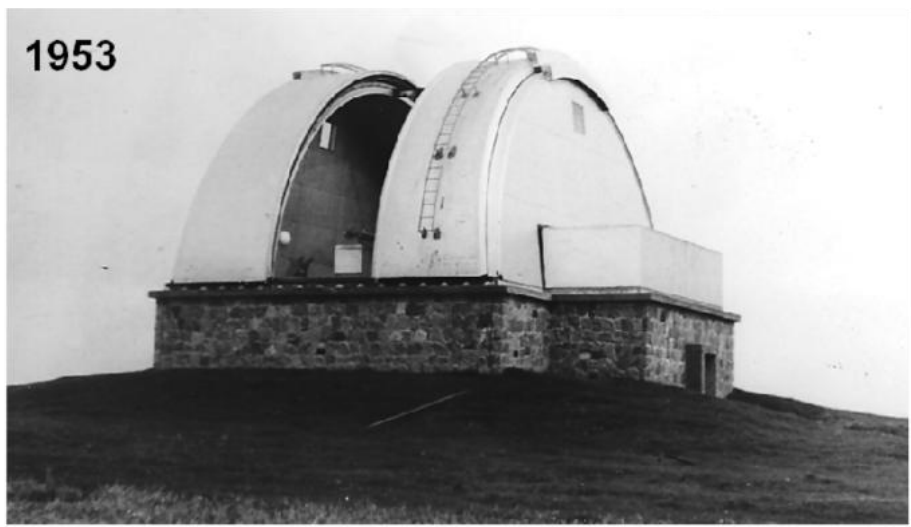

Figure 7. The observatory at Østervold built in 1861, Tycho Brahe in the front.
– The meridian pavilion in Brorfelde built in 1953 – Photos: Erik Høg

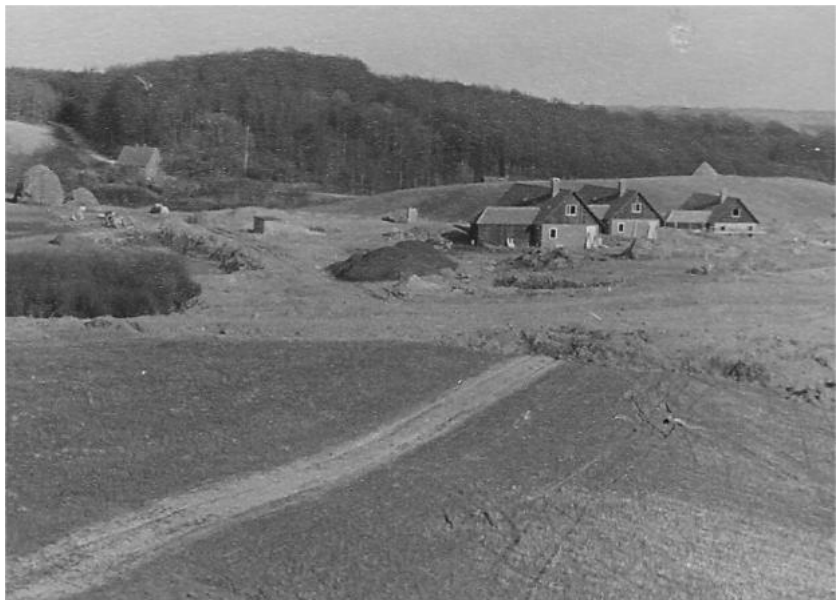

Figure 8. View from the top of the meridian pavilion towards east in early 1955. Three houses for personnel are seen, a fourth and fifth closer by came later. The road along the houses is still missing. The main building is further away in the same direction, but hidden behind a hill. Even further appears the top of the mounting tower of the workshop. – Photo: Erik Høg

The fifth staff member of the observatory, Kjeld Gyldenkerne, was stationed near the small town of Tølløse since September 1945, at the railway from Copenhagen and near the coming Brorfelde observatory location. Gyldenkerne, then newly married to Inger, came to stay in the rented house in Gammel Tølløse for 12 years, much longer than expected because of the slow progress with the observatory in Brorfelde. He and his family finally moved to the fourth house in June 1957. Inger Gyldenkerne (*1921) told me all this on the phone in January 2015. During those years Gyldenkerne carried out photoelectric photometry as initiated by Strömgren for the development of an accurate colour filter system for galactic research, soon known as the Strömgren four-colour photometry.

Finally, the sixth scientific figure in astronomy in our part of Denmark must be mentioned, Einar Hertzsprung (1873-1967). After retiring as director of the Leiden Observatory in 1946, he moved to a house in Tølløse near the railway station where he stayed until he died in 1967 at the age of 94. He had chosen Tølløse to enjoy the scientific life in the observatory then expected soon to come in Brorfelde. I often visited Hertzsprung in Tølløse and later corresponded with him, the last visit was when I came with my wife Aase shortly after we had married in December 1962. He lived there with his daughter, and as usual she served a fruit dessert for the three of us in the dining room to end the visit.

Erna Mackeprang was the secretary at Østervold. She worked much with Vinter Hansen on the yearly University Calendar which was edited and published by the observatory. Man-of-all-work was Jensen. "Our good old Jensen" as Julie called him was living with his wife in a flat in the east wing of the observatory where also Vinter Hansen had her apartment, and the library was located there. The entire west wing of the building was occupied by the large apartment and the splendid office of the professor, all empty after Bengt Strömgren and family had left in 1951.

In 1954 Naur and I moved into one of these fine rooms. He was my senior by only four years, but had already much experience when he became my mentor in October 1953 after returning from USA and England. During the following two years I shared the spacious office at Østervold with Naur until he moved to Brorfelde. I had of course my own desk and could discuss with him any time, so he was my window to astronomy and to the then upcoming world of computers.

Astronomisk Selskab, the Astronomical Society, held regular meetings where a lecture was given, usually in Danish, followed by a social gathering in a restaurant near the observatory. I mostly participated and find e.g. a note in my calendar under 5 October 1951 that Naur spoke on electronic computers. I do not remember that talk but Naur was always a brilliant speaker.

### 3.3 My "Big Task" is defined

A student must learn his science at lectures and from reading and discussions and he must show his ability for scientific work in a big task, defined by his teachers. In my case this process can be followed in some detail: I did the scientific work very well, according to the professors Reiz and Strömgren whom I had informed by letters, but my final written report in 1955 consisting mostly of these letters was far from satisfactory in its form, as I shall explain in the following sections.

It appears from letters between Vinter Hansen and Strömgren during two months, November and December 1953, that the work for me was thoroughly discussed, also with Naur. It was in those days called "stor opgave" in Danish, i.e. "Big Task" and had to be completed with a report after one or two years. I had never seen how such a report should look and this caused some turmoil. I will quote extensively from these letters since it is perhaps seldom documented in such detail how three supervisors of a young student discuss what is the best procedure; and I am very happy to see how considerate and thoughtful they were in my case. In reality, they did not have much other choice than to select me because I was the only student in question and the meridian circle required immediate attention to justify the big investment to the sponsors. It was however a matter of good luck that this only student had the talent and enthusiasm to work meticulously under "very unpleasant conditions" as one of the supervisors, Peter Naur, meant.

The supervisors discussed whether I was good enough, and I want to confess now that I have not always been sufficiently careful and meticulous. One day, probably in 1958, Gyldenkerne asked me to set up the north-south line for the concrete foundation of the Schmidt telescope. I did so by measuring directions to distant identified points in the landscape, using the proper instruments. But Gyldenkerne later told me, with a smile I think, that my line was wrong by several degrees. He had checked it and he could make no mistake with his experience as a land surveyor. It would have been a shame if they had used my line to pour the concrete. I had made a fool of myself and I shall not forget.

The discussion begins on 3 Nov. 1953 when Vinter Hansen writes to Strömgren that Naur is thinking that he and Høg should determine the division corrections together. Høg shall make the exposures and Naur will have the measures punched. But Vinter Hansen had understood that Strömgren intended to give Høg a more independent task.

In Sect. 5.2, I explain division corrections and their importance for accurate observations and that it took 14 years before all the division corrections were successfully determined in Brorfelde. A big step towards a realistic solution was taken five years later in Brorfelde when I invented a new method, published as Høg (1961). A big task it was, larger than anybody had imagined. Is it therefore an example of poor planning? Well, such is real life sometimes, also in science. What matters most is the result in the end: success or failure.

On 12 Dec. Strömgren writes that Høg should work with the meridian circle for about half a year in one stretch. He should measure collimation, nadir and inclination 3-4 times in 24 hours to get a good idea of the instrument stability. Furthermore he should determine the azimuth error by transit observations as Reiz, Rudkjøbing and Naur had done in October. All that would be useful and it would be instructive for Høg. But after half a year Høg should have a task of less routine character.

About the division corrections, Strömgren thinks that Høg can at most be an assistant to Naur. It is such an important task that not even taking the photographic exposures should be left to Høg alone. "In reality I have my doubt whether he will be careful enough for such a task, but I believe it would be OK, and his results will show whether he does it well or not. – However, the development may show that it could be more useful if he stays in Copenhagen for a while and gets some other work."

On 18 Dec. Vinter Hansen writes that she has spoken with Naur and he thinks that as long as there is no place to live near the meridian circle it would be rather damaging for the interest and energy of a young man if he should work under the present very unpleasant conditions uninterrupted for half a year just measuring collimation, nadir and inclination. Vinter Hansen is also in doubt whether Høg is sufficiently careful for such important measurements. It would therefore perhaps be better to give him tasks in Copenhagen, but which?

On 2 Jan. 1954 Strömgren writes: "Høg must of course first make useful work in Copenhagen with some routine job where he is sitting every day in the observatory, and does work of a kind where control and spot testing is easily done. Later on he can work with something more exciting". – Strömgren then mentions computations related to Naur's work on Nemausa. The minor planet 51 Nemausa moved fairly close to the celestial equator and its positions as observed by photographic astrometry were therefore used to determine the systematic errors of the system of declinations, systematic errors which had been bothering astronomers for many years.

On 7 Jan. 54: Vinter Hansen has spoken with Naur and he will put Høg to work. - In fact I remember doing e.g. interpolations in the tables, Naur had produced in Cambridge using the electronic computer EDSAC.

After this date and during more than two years I find nothing about Høg in the letters.

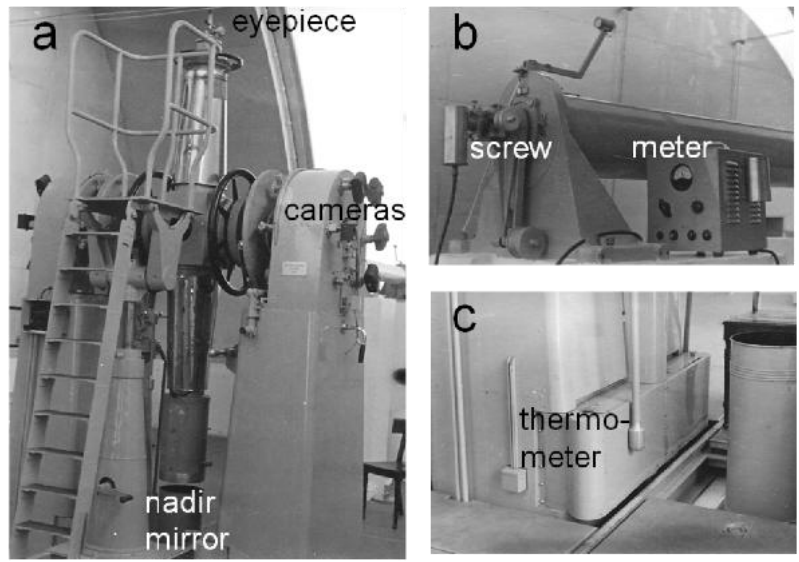

Figure 9. ***a:*** The new meridian circle in Brorfelde in 1955. After climbing the ladder, the nadir direction could be observed by means of a visual Bohnenberger-eyepiece. Illuminated wires in the eyepiece were reflected by the horizontal surface of a mercury pool below. - Photographic cameras (black) for recording the circle are seen at left and right. They had been delivered by Grubb Parsons together with the meridian circle, but they were completely unfit for use and other cameras were built in the mechanical workshop. ***b:*** Photoelectric device to align the collimators. After alignment the slits in the collimators were photographed with the main telescope. ***c:*** One of six thermometers with 0.1 deg division mounted on the pier. – Photos: Erik Høg

**3.4 Meridian circle stability**

During almost one year, August 1954 to July 1955, I worked much in Brorfelde testing the stability of the instrument. But all the time, in fact 1 April 1953 to 15 Feb. 1958, I lived at Elers Kollegium, one of the oldest student dormitories near the Round Tower in Copenhagen. I had a room and later even two rooms enjoying the life among students. I could drive the 50 km to Brorfelde on my small motor bike, a NSU 125 cc, which I had bought fairly cheaply with special permission because of my work at night in Brorfelde. Cars and motor bikes were very expensive in Denmark in those years.

Access to the meridian pavilion was sometimes difficult because there was no paved road yet. Figure 8 shows a road from the lower left corner of the picture ending in the middle of the picture where a mud pool would form in rainy weather. I once got stuck in the mud in my rubber boots – next time I arrived from the main road behind me where I left my motor bike and walked up the hill to the pavilion.

Since there was no other building than the meridian pavilion I was completely alone. I made visual observations of collimation, nadir and inclination and both visual and photographic observations of polarissima, a star very close to the celestial pole (named BP+89.3). I sometimes used a photoelectric arrangement designed by Naur to align the collimators, Figure 9.a. When clouds prevented further observations I sometimes slept in a haystack or on the floor in the small dark room where the photographic plates were developed and which could therefore be heated.

I measured the plates in Copenhagen and reported by letter to Strömgren, first on 5 and 6 Oct. 1954, four pages in total. Strömgren answered on 16 Oct. with two full pages in his nice and very readable hand. He was very satisfied and gave good advice back. I lent copies to Anders Reiz, who also replied and thanked. A seven-page report of 21 May 1955 was sent to Strömgren and to Reiz and they replied positively.

My Big Task report is dated November 1955 and was handed to Vinter Hansen. It has the title (translated) *"Studies of the stability of the meridian circle in Brorfelde August 1954 to July 1955"*. It contains 20 typewritten pages beginning with a summary. The summary ends thus: "The report consists of 20 pages plus the material behind as to be found in protocol M1, the notebooks M1a, M1b and the loose pages Mø by means of notes in the margins. Included are also 13 photographic plates."

On 11 March 1956 Strömgren writes: "I think that Høg's work at the meridian circle was excellent, also the report he sent. I assume it has become a satisfactory Big Task. Has Høg completed the examinations? Then he probably has to do his military service." –But a serious disaster threatened!

Before I tell about the near-disaster I would like to give an example of Julie Vinter Hansen's great sense of humour. In a long letter on 24 May 1955 she writes to Dear Bengt! … "Our great concern is our good old Jensen who is still ill … he has been replaced by Reidl. … Reidl is the fireman who some years ago literally broke his neck and stayed in hospital for a couple of years, so he is no strapping fellow but he is busy as a beaver and agreeable. He is distinguished by having all visible parts of his body, except the face, very nicely tattooed. I have asked him if he has a lovely lady on his chest, and he has; but I have not seen her." – Julie was a gifted story-teller and I could fill pages with her anecdotes, mostly funny ones, from a long life among astronomers at Østervold and elsewhere e.g. Elis Strömgren and E. Hertzsprung, and the older generation at Østervold T.N. Thiele and C.F. Pechüle.

**3.5 Disaster averted by Julie Vinter Hansen**

I have hesitated to include the following because it is far from flattering for myself, but it is rather amusing in hindsight and it illuminates those times in the Copenhagen observatory.

On 27 March 1956 Vinter Hansen writes to Strömgren: "Regrettable it has not gone as well with Høg's Big Task as I had expected. His observations and measurements are fine, but he has been very sloppy in presenting his procedures. A few months ago I gave him the report back and told him that he ought to rework the various parts into a more coherent text. When he gave it back I spent much time to penetrate the mysteries and became quite sad at last, because I believed I had become senile. Then I gave it to Naur without a comment and his verdict was very tough; I believe that he would have preferred to give it back again for a new rework. But I thought that would be too tough because the observations had been done well. We, i.e. Rudkjøbing, Naur and I, then agreed to call him in and tell him that we would let his Big Task pass, but that his exam would depend on how he did the two 4-hour written tasks and the oral examinations. He was called to become a soldier in May and it would also for this reason be disastrous for him if the report would not be accepted. Naur admits that he has probably spoiled him a bit. However, I cannot believe there is any danger that Høg will not pass, but the situation is annoying and for me quite unexpected."

The oral examinations on 12 April were indeed quite tough, lasting about three hours I think, but I did pass the final exam. A week was given to prepare the public lecture on 19 April with the title, translated: *"The differential rotation of the Milky Way and the evidence provided by meridian circle observations."* This was the "Magisterkonferens" allowing me to carry the title of cand. mag. (candidatus magisterii) so that I could teach in grammar school and the title mag. scient. (magister scientiarum), Master of Science, which qualified me for research. Altogether in the 14 years from 1949, six persons passed the main exam in astronomy in Copenhagen: Peter Naur in 1949, Heimann Ever Jørgensen (*1926) in 1954, Erik Høg and Svend Laustsen in 1956, Jørgen Otzen Petersen (*1935) in 1960, and Henning Elo Jørgensen (1938-2010) in 1963. Vagn Mejdahl (1928-1997) should be mentioned here because he worked one year on the meridian circle in Brorfelde from 1 July 1959, but his exam in 1957 was in physics. The study of astronomy became possible also at the Aarhus University when Mogens Rudkjøbing on 1 April 1956 was nominated as professor in astronomy.

Subsequently I bought a book about the writing of a technical report and studied it intensely, improving gradually. I completely agree on the verdict of my report, I had not given sufficient attention to writing a coherent text. I have since met many colleagues with a born talent for writing coherently, but I had to learn the hard way. Furthermore, I never learnt to use ten-finger typing although I tried hard when I was young and I immensely admired Naur when I saw him using all fingers. Today, with a text editor retyping and rearranging is infinitely much easier than in the old days. We wrote, then cut out with scissors and glued together, and finally retyped the whole thing. Few of us had a secretary.

# 4  Soldier in a laboratory

> During the time as a conscript soldier I worked with electronic counting techniques in a laboratory. This technique was quite new at the time and it became very important for my ideas of photon counting astrometry (see Figure 5) a few years later. As so often, I must say that I have been at the right place at the right time meeting the right people: for instance soon after in Brorfelde, and after 1958 at the Hamburg observatory as I have told in Høg (2014a).

### 4.1  Military service 1956-57

After the final exam, I entered the obligatory military service period of 16 months in May. It consisted first in three basic months training in the field, but continued in my case with work in a radio-physical laboratory in Copenhagen. We collected rain and dust at several places and dry samples were prepared for measurement of the decay of radioactivity in fallout from the ongoing atomic bomb testing of the superpowers in the atmosphere. This work made me familiar with electronic counting techniques, something very new at the time and it became very important for my ideas of photon counting astrometry a few years later.

On behalf of the radio-physical laboratory, I attended a geophysical conference in Utrecht in Holland in 1957 and another in Moscow in 1958 because the laboratory was contributing to the international geophysical year 1957/58. The 10-day conference in Moscow was followed by the General Assembly of the International Astronomical Union, also of 10 days. The three weeks in Moscow gave opportunity to many activities, only one of which I shall mention here. I had been asked by a commander from the Danish military intelligence service to inquire about electronic computers in the USSR if possible. He visited me in Brorfelde to explain what he wanted.

When I asked about computers at the conference centre in Moscow, the answer came several days later: "The director on this matter was on vacation." But I did not need to see the director, I said, I just wanted to talk with someone who knew about the computers. In the end, a group of astronomers was invited to see the computers and ask questions. I was not impressed. What we saw was not on a level with the Danish computer DASK produced by Regnecentralen and completed in February 1958. But the Russians did probably not show us the best they had because in that case they would hardly have been able to launch the first artificial satellite Sputnik a year before. Back in Denmark, I wrote a report to the commander as I had promised.

In 1957, together with Svend Laustsen I made visual observations of the comet Arend-Roland with the refractor in the dome at Østervold while I was a soldier in a laboratory, and we calculated an orbit from three observations. Laustsen's Big Task had been on the motion of the asteroid 51 Nemausa based on photographic astrometry. After his exam in 1956, Laustsen stayed five months in Leiden in 1957. He then worked at Østervold and Brorfelde with the meridian circle until 1968 as described below. He worked at ESO 1968-78 and then at the Institute of Astronomy in Aarhus until his retirement in 1990.

# 5  In Brorfelde 1957-58

After the military service, I started in Brorfelde from October 1957 working with Peter Naur to make the meridian circle operational. The splendid main building was completed and I had a room with bed and a shower, next to the kitchen and the living room. The next year I had to cook for myself, something I had to learn, and I could invite guests. After working at Østervold I could still stay overnight at Elers Kollegium, in Tullianum.

### 5.1  Electronics

My office was next to Naur's and we used the basement under the library as electronics lab. To begin with I learnt practical electronics under Naur's guidance, and this became very important soon after when I came to the Hamburg observatory where nobody else could do that. In Brorfelde I built RC-oscillators with 12 and 23 watts and wrote a report with seven pages, dated January 1958. I presented it to Vinter Hansen when she visited us in the Brorfelde lab and she was delighted. The report describes the electronics and results of testing. It also details the work involved: The components for the 12 watts oscillator had cost about 350 kroner, I had spent 53 hours on planning and mounting, 57 hours on experimentation, 15 hours on drawings and the report, in total 125 hours. Peter Naur had spent 10 hours giving advice because I was new in the field. The work was done from 31 Oct. to 6 Dec. 1957. The oscillator however was too weak to drive the 50 cm Cassegrain reflector for which it was planned. The 23 watts was therefore built and the work is also specified in the report. – Julie, of course, did not make any hint to my problematic Big Task report two years earlier. I really admire the good old lady for her kindness and for her hard administrative work during the many years when the professor was absent from Copenhagen most of the time.

Naur worked on several issues concerning the meridian circle; he described the Watts machine in a publication from the observatory in 1958; he studied the problems of the clocks to be used, especially the use of the then quite new transistors; and he wrote an 8-page note in 1958 on "A transistorized quartz oscillator". I remember very well that Naur presented the transistor, a small black thing, to a few of us as a revolution in electronics. I tried to understand the transistor but never well enough to design electronics with transistors, I continued with radio valves which was still good for some time.

**5.2 Division lines**

Naur and I had divided the work on the meridian circle such that he worked on the measurement of right ascensions and I took care of the declinations. A machine to measure films from the declination cameras (Figure 9.a) had been developed at the US Naval Observatory and this "Watts machine" had been copied in our workshop. Under Naur's guidance I started to build the electronics from drawings received from the USA. The films showed the division lines from the declination circle imaged by microscopes attached to the pier of the meridian circle. The measurement would give the inclination of the telescope at each star observation, and thus the declination of the star.

I will now explain in some detail how a formerly gigantic task in meridian astronomy was converted into something manageable and much more accurate. - The division lines were very nicely etched on the glass circle with 12 lines per degree, but they were not placed accurately enough. They needed calibration because it was the very purpose of a meridian circle to measure large arcs on the sky very accurately, i.e. declinations along the meridian, and right ascensions by timing the transit times of the stars.

It was my task to determine the division corrections for the several thousand lines, so I studied the literature on this subject, and realized the enormity of the task. With the best method described so far (Levy 1955), it would require some 50 000 measures, at least 1000 man-hours with visual reading of the microscopes, and in reality much longer. Previously it had taken astronomers several years to do the whole job. I saw that the arrangement of the microscopes had to be changed many times, since the corrections could then be calculated from the measures simply by additions and subtractions, no multiplications were needed, and multiplications were out of question in those days.

Soon however I could show that the use of six microscopes in two different positions instead of the usual four microscopes in many different positions, would greatly decrease the number of required measures, from 50 000 to 15 000, and improve the accuracy. This assumed what nobody had considered before, that a least-squares method was applied as was becoming feasible with electronic computers.

When I told Naur about the idea, he became very interested and arrived the next morning with more ideas. I did not like that and said: "You are much brighter than me, and nothing interesting will be left for me!" He immediate understood and we continued our collaboration where declinations remained my field. Shortly afterwards, we visited the DASK computing centre in order to get a solution of the resulting normal equations which Naur had in fact helped to prepare. Two systems of linear equations with 11 unknowns were quickly done with DASK but would have been a big job to do with our usual hand calculators. On 13 February1958, a week after our visit, we received the two solutions and a bill of 45 Kroner (=6 Euro?). - My work on determination of division corrections was completed soon after in Hamburg, see Sect. 6.2.

### 5.3 Automatic measurement

During the time in Brorfelde, in April 1958, I described my vision about automatic measurement of photographic plates by complete scanning of the plate and storing the data on magnetic tape for subsequent treatment in an electronic computer. The report (Høg 1958) was distributed to about 30 persons and was well received. Peter Naur, e.g., sent it to professor J.H. Oort in Leiden with an accompanying letter of 6 May 1958 where he describes the proposal "by our young assistant here, Mr. Erik Høg" for using "the modern electronic computers for relieving us of all the troublesome measurement of plates…". Subsequently, I received a kind letter from Oort.

Obviously, I was thriving in Brorfelde, but Peter Naur encouraged me to go abroad for some time. I objected that I did not know enough, I had to learn more before going. But he insisted saying I should go now, precisely in order to learn more: "Out there you will learn". He wrote a letter of recommendation of 23 May 1958 where he praised my work on the meridian circle and on automatic measurement of plates. His letter ends with the following, translated from Danish: *"Regrettably, his education has taken place at an unfortunate time for the teaching of astronomy at the Copenhagen University, and it is therefore especially desirable to give him the opportunity for further qualification during a stay abroad under the best possible circumstances."*

### 5.4 Peter Naur leaves astronomy

Since I was quite happy with the work in Brorfelde, I did not quite understand Naur in those days when he sometimes said that the leadership of the observatory was unsatisfactory, even incompetent. But I can now see that the circumstances were scientifically poor and frustrating for Peter Naur who was missing Bengt Strömgren as the natural and competent leader of the observatory; Naur told me recently that he did not have any talks with Strömgren during the two years 1957-58.

Naur had stayed abroad. At Cambridge in England he used the electronic computer EDSAC and he learnt to build electronics. At Yerkes and McDonald observatories in the USA he had been observing and he had felt the great scientific atmosphere.

He had already been a regular guest at the Østervold observatory while he was a schoolboy and while Strömgren was present. Naur had e.g. calculated orbits of comets. It was therefore quite natural, but a great surprise for me when I heard that he left the observatory and astronomy entirely in January 1959. In Section 10 of Høg (2014a) I tell about the year 1957-58 as his assistant in the Brorfelde observatory where I learnt to design and build electronics and I tell about Naur's significant role in astronomy and computer sciences. Naur calls it one of the happiest choices of his life when he succeeded to leave the Copenhagen Observatory in 1959 and start a career in computer sciences, or "datalogy" as he prefers to call it.

# 6  The revolution of astrometry

In Hamburg from 1958, I began to work on astrophysics with plates from the big Schmidt telescope. I wanted to get away from astrometry since this by everybody was considered to be an old fashioned branch of astronomy. But in July 1960 I got the idea of photon counting astrometry and was irresistibly drawn into astrometry again as I have told in Høg (2014a). After my talk in 2012 in the Astronomische Gesellschaft on this matter and the Hipparcos mission, a comment was this: *"You said you wanted to get into astrophysics fifty years ago, but by returning to revolutionize astrometry you have done more for astrophysics than you could have done as an astrophysicist"*.

### 6.1  Photon counting astrometry and new satellite design

In the Hamburg-Bergedorf observatory, I developed photon counting astrometry for the meridian circle expedition (1967-73) to Perth in Western Australia. This was the fruit of a seed laid by Peter Naur when he drew my attention to the experiment by Bengt Strömgren in 1925 with photoelectric recording of stellar transits across slits, cf. Figure 5. This seed had grown and ripened in my mind in July 1960, see Høg (2014a), and the director professor Otto Heckmann (1901-83) immediately saw the potential.

The scheme of photon counting astrometry was appreciated especially in France where the director of Strasbourg Observatory professor Pierre Lacroute (1906-93) in 1964 began to develop his idea of space astrometry, using this technique and a special mirror telescope of his design. In 1975, the European Space Agency (ESA) formed a mission study group on space astrometry to which I was invited. In December of the same year I presented a new more realistic design of the satellite and mission, and the resulting study report led to thorough studies by ESA and to the approval in 1980 of the astrometry satellite Hipparcos which revolutionized astrometry. The astrometric satellite would not have been approved in 1980 and probably never, if anyone of seven named persons had been missing, according to Høg (2011). Bengt Strömgren is the first with his experiment in 1925 on photoelectric astrometry and I am another of the seven persons.

Returned to Denmark in 1973, Hipparcos soon became my highest priority. But I had to work on Hipparcos with external funding and I was lucky to find good collaborators. My observatory did not take any responsibility for the participation in space astrometry and all institute resources for astrometry were given to the work on the meridian circle.

## 6.2 Division lines

In Hamburg I completed the work from Brorfelde on determination of division corrections. I realized that the system of normal equations could be solved explicitly by trigonometric functions because of the cyclic structure. The "general symmetric method" was published as Høg (1961) and it was applied in Hamburg-Bergedorf by von Fischer-Treuenfeld (1968). The method has been used ever since by meridian astronomers in various versions.

Laustsen and collaborators in Brorfelde at first used new photographic cameras to record the circle, replacing those delivered by Grubb Parsons (Figure 9.a). The films were measured with the Watts machine from 1958-66, according to Sect. 3 of Einicke et al. (1971), but without good results, according to Laustsen's memory. Therefore, from December 1965 a photoelectric scanning method was applied. Two years later, this technique had been refined to the point that all observations and the basic reduction, including division corrections, could be done in less than 24 hours (Einicke et al. 1971).

The photoelectric scanning method had been proposed in 1960 in Sect. 4.2 by Høg (1960) and was also implemented in Hamburg. Another reason for the success was that a GIER computer, the transistorized successor of the Danish DASK, had been installed at both observatories. GIER was ten times faster than the best other affordable computer on the market and that was decisive for our work, see Sect. 5 of Høg (2014a). - A decade later, Miyamoto & Kuehne (1982) obtained all the division corrections with the automatic meridian circle in Tokyo in less than one day, and repeating this day after day, time variations could be observed.

## 6.3 Historians and history of astronomy

The work by Gyldenkerne et al. (1990) on Danish astronomy consists of three volumes in *de luxe* binding. Volume 2 of 310 pages in Danish contains a chapter 37 about Danish astronomers abroad. Kaj Aage Strand (director of the US Naval Observatory in Washington DC) is here mentioned on half a page and I remember clearly his dissatisfaction when I met him in about 1991. This came to mind recently when Peter Naur complained to me that none of us is mentioned anywhere. Gyldenkerne does mention the photoelectric technique on p.288, furthermore on p.305 the "satellite HIPPARCOS-TYCHO" and also my glass meridian circle experiments, but no person is named.

Gyldenkerne completed his book just after Hipparcos was launched in 1989, and he knew very well my photon counting astrometry from July 1960 and my new realistic design of the satellite mission in December 1975. He knew that my work had brought the revolution of astrometry which was bound to come with Hipparcos, cf. Sect.6.1. He also knew that I proposed Tycho in 1981.

With Svend Laustsen's permission, I can report that he was asked in 1987 by the publishing firm Rhodos to become editor of the three volumes on Danish astronomy, but he declined, and Gyldenkerne then took the task. Laustsen adds that he has often wondered about the considerable shortcomings in the description of the history of Danish astronomy in the 20[th] century, but in spite of this we both admire the work of Gyldenkerne as editor at the age of 69 in 1990. He would have faced an immense task if he had tried to do justice to all excellent scientific work at the observatory.

When asked by Michael Perryman for his book in 2010 on the Hipparcos mission which awards I had received for my work from my own country, I had to answer that I had received none. In the book on p.262 are listed the awards received by the four leaders of the Hipparcos and Tycho consortia, for me the Struve Medal I received in Saint Petersburg in 2009 and for the other three their national awards. Soon after with NBI (2010), my work was placed prominently at the website of the Niels Bohr Institute.

The historian, here Gyldenkerne, would enter a minefield if he would mention any other outstanding persons than the established heros, here Strömgren and Hertzsprung, there would be too many to include. I noticed a similar tendency at the centenary of the Hamburg-Bergedorf observatory in 2012. The heros in Bergedorf are especially Bernhard Schmidt and Walter Baade and the astrometry comes second. The meridian circle expedition to Perth was not mentioned at all by the astrophysicists at official occasions, and it was and still is (February 2015) neglected on the Hamburg observatory website.

But a person who takes responsibility for the history of science is professor Gudrun Wolfschmidt (Center for History of Science and Technology, Hamburg Observatory, Department of Physics, Faculty for Mathematics, University of Hamburg). She had invited me to lecture at the centenary and I wrote Høg (2014a) in which the Sections 11 and 12 discuss historical issues about astrometry in respectively Hamburg-Bergedorf and the European Space Agency.

# 7  Fifty years Brorfelde

> The 1960s must be mentioned for the beginning of great developments in Danish astronomy with a large staff working in astrophysics, astrometry and instrumentation. The new technique, photon counting astrometry developed for the meridian circle in Hamburg was needed in Brorfelde and I was called back to Denmark in 1973, bringing my experience. This led to the development of the Carlsberg Automatic Meridian Circle, operating on La Palma since 1984. Most importantly, the photon counting technique was required for the ESA astrometry satellite Hipparcos launched in 1989, opening a new era for astrometry, an era with roots in Brorfelde.

### 7.1  Østervold and Brorfelde 1959-96

The golden 1960s allowed professor Reiz to expand the staff and develop an instrument park, and Denmark joined the European Southern Observatory in 1967 as the sixth member. In 1964, Svend Laustsen was able to make the first useful observations with a photographic camera on the meridian circle in Brorfelde. By 1986, the staff in Brorfelde counted 27 employees including 9 astronomers, see Gyldenkerne (1986). Instruments poured out of the mechanics and electronics workshops, both uniquely efficient under the leadership of respectively Poul Bechmann (1916-2000) and Ralph Florentin Nielsen (1940-1995).

The era of astronomy at the Østervold and Brorfelde observatories ended when the staff from both places moved to a new location in Copenhagen, fifty years after Brorfelde had been chosen.

In early 1996 they, including myself, settled in the Rockefeller Complex at Juliane Maries Vej 30, in Copenhagen together with geophysicists and the Danish Space Research Institute. Before the move, the archives at the Østervold observatory were registered at the initiative of the director professor Henning E. Jørgensen (1938-2010) who began as a young student of astronomy in September 1956. The registrant and papers were then ready to be sent to an archive, according to recent information from Claus Thykier (*1939), founder and leader of the Ole Rømer Museum, who supported this work of registration.

This section should close with an emphasis on the significant role of the meridian circle in Brorfelde in spite of the first many years with slow progress – as it seemed then for those involved. The experience gained in Hamburg-Bergedorf in the 1960s sprang from the work in Brorfelde in the 1950s and I returned to Brorfelde in 1973. Merging these experiences in Brorfelde in the 1970s, the meridian circle was further developed. Led by Leif Helmer (1943-) and supported by British and Spanish partners, the instrument was moved to La Palma in 1982 and began operation in May 1984 as the Carlsberg Automatic Meridian Circle (see Section 3 and Table 1 in Section 9 of Høg 2014a). It became the most efficient meridian circle ever and stands as a splendid completion of the 300 year era of meridian circles.

Space astrometry has taken over the accurate measurement of large arcs between stars with the two ESA astrometry satellites Hipparcos and Gaia launched in respectively 1989 and 2013.

**Acknowledgements:** I am grateful to Gudrun Wolfschmidt for the invitation to contribute to the meeting "Astronomie im Ostseeraum" and to Lars Brink, Mia Bjørg Olsen, Claus Fabricius, Inger Gyldenkerne, Thomas Gyldenkerne, Leif Helmer, Ole Henningsen, Aase Høg, Povl Høyer, Hans Henrik Larsen, Svend Laustsen, Javier Montojo Salazar, Peter Naur, Karen Florentin Nielsen, Birgitte Pantmann, Holger Pedersen, Jørgen Otzen Petersen, and Claus Thykier for discussion and information. Finally I want to thank two persons for urging and encouraging me since 2009 to write my scientific biography: Rajesh Kochhar (President International Astronomical Union Commission 41: History of Astronomy) and Gertie Skaarup (Editor in Chief, Communication Department, Niels Bohr Institute). – I am doing my best, but it comes piecewise.

# 8  References

The reader of a printed version should note that the present report is placed on my website so that the internet links to most references are directly available through
http://www.astro.ku.dk/~erik/xx/BalticDan1.pdf  This link replaces in 2017 the now invalid link to my dropbox.

Andersen, M.C. 2004, Astronomi ved Københavns Universitet 1957-2003. In: *Almanak* 2004, 118-137, Københavns Universitet

Einicke, O.H., Laustsen, S., Schnedler Nielsen, H. 1971, Precision of Circle Reading and Determination of Diameter Corrections. Astronomy and Astrophysics, Vol. 10, p. 8 (1971).
http://esoads.eso.org/abs/1971A%26A....10....8E

von Fischer-Treuenfeld, W.F. 1968, Die Anwendung der allgemeinen symmetrischen Methode bei der Teilkreisfehlerbestimmung. Dissertation: Hamburg, Math.-Nat. Fak., Diss. v. 8. April 1968. Verlag: Bönecke, Clausthal-Zellerfeld, 1968

González, F.J.G., 2004, El Observatorio de San Fernando en el siglo XX. ISBN: 84-9781-091-0, see pages 166 – 170 in the third Volume

Gyldenkerne, K. 1986, Fyrre år i Brorfelde (Forty years in Brorfelde). Nordisk Astronomisk Tidsskrift Årg. 19, Nr. 3, p. 97-119 in Danish with 23 figures, including photos of buildings and instruments and of the 27 people employed in Brorfelde in 1986. http://esoads.eso.org/abs/1986ATi....19...97G

Gyldenkerne, K., Darnell, P.B. et al. 1990, Dansk astronomi gennem fire hundrede år (Danish astronomy during four hundred years). Editor: C. Thykier. Rhodos, internationalt forlag for videnskab og kunst. Three volumes. An English summary is contained in the last volume on pp.583-589

Høg, E. 1958, Automatic measurement especially of photographic plates using the electronic computer. Internal report, dated April 20, 1958, typed and copied in the observatory at Østervold, unpublished. 30 pages

Høg E. 1960, Proposal for a photoelectric meridian circle and a new principle of instrumental design. Astron. Abh. der Hamburger Sternwarte, vol. V, 263-272. At: http://adsabs.harvard.edu/abs/1960AAHam...5..263H

Høg, E. 1961, Determination of Division Corrections. Astronomische Nachrichten, volume 286, p.65. http://esoads.eso.org/abs/1961AN....286...65H

Høg E. 2011, Astrometry lost and regained. In: Proceedings of the international conference "Expanding the Universe", Tartu, Estonia 2011 April 27-29 on the occasion the 200th anniversary of the Tartu Observatory. Edited by C. Sterken, L. Leedjärv and E. Tempel. *Baltic Astronomy, vol. 20, 221,* and at www.astro.ku.dk/~erik/BalticAhoeg.pdf

Høg, E. 2014a, Astrometry 1960-80: from Hamburg to Hipparcos. Proceedings of conference held in Hamburg in 2012, Nuncius Hamburgensis, Beiträge zur Geschichte der Naturwissenschaften, Band 24, 2014. http://arxiv.org/abs/1408.2407

Høg, E. 2014b, The Astrometric Foundation of Astrophysics. Abstract to the Conference Book 2014 of the Danish Astronomical Society and abstract of a review presentation. http://arxiv.org/abs/1408.2122

Levy, Jacques 1955, Détermination des Corrections de Graduation des Cercles Divisés. Bulletin Astronomique, vol. 20, pp.35-78. http://esoads.eso.org/abs/1955BuAst..20...35L

Miyamoto, M.; Kuehne, C. 1982, An accurate derivation of the division corrections in a photoelectric meridian circle. Astronomy and Astrophysics Supplement Series, vol. 50, Nov. 1982, p. 173-186. http://esoads.eso.org/abs/1982A%26AS...50..173M

Moesgaard, K.P., Pedersen, K.P., Strömgren, B. 1983, Astronomi. In: Københavns Universitet 1479-1979, Vol. XII, 247-363.

Moustgaard L. 1990, Uranias tjenere - Episoder i dansk astronomi 1900-1950. København 1990. 159 pages.

NBI 2010, Niels Bohr Institute - Who, What, When. At http://www.nbi.ku.dk/english/www/

Nielsen, A.V. 1962, Hundrede års astronomi på Østervold (Four hundred years astronomy at Østervold). Reprint of Nordisk Astronomisk Tidsskrift 1961-62

Perryman, M. 2010, The Making of History's Greatest Star Map. Springer-Verlag Berlin Heidelberg, 282pp.

Petersen, J.O. 2015, Observatoriet på Østervold i vækstperioden 1958-1975, DASK og GIER – dengang og nu. In: KVANT, maj 2015, 26-35